\documentclass[twocolumn,floatfix,aps,prd,superscriptaddress,nofootinbib]{revtex4-2}

\usepackage{iftex}
\ifPDFTeX
  \usepackage[utf8]{inputenc}
  \usepackage[T1]{fontenc}
  \usepackage{lmodern}
  \input{glyphtounicode}
\else
  \usepackage{fontspec}
\fi
\usepackage{graphicx}
\usepackage{amssymb}
\usepackage{amsmath}
\usepackage{xcolor}
\usepackage{hyperref}
\usepackage{placeins}
\usepackage[protrusion=true,expansion=false]{microtype}
\usepackage{tikz}
\usetikzlibrary{positioning,arrows.meta}
\definecolor{figink}{HTML}{1A1A18}
\definecolor{figmuted}{HTML}{5F5D57}

\AtBeginDocument{%
  \setcounter{dbltopnumber}{2}%
}

\makeatletter
\newcommand{\packfloatpages}{%
  \setlength{\@fptop}{0pt}%
  \setlength{\@fpsep}{12pt}%
  \setlength{\@fpbot}{0pt plus 1fil}%
  \setlength{\@dblfptop}{0pt}%
  \setlength{\@dblfpsep}{12pt}%
  \setlength{\@dblfpbot}{0pt plus 1fil}%
}
\makeatother

\graphicspath{{figures/}}

\hypersetup{
    colorlinks=true,
    linkcolor=blue,
    filecolor=magenta,
    urlcolor=blue,
    citecolor=blue,
}
\begin{document}

\raggedbottom

\title{Rotation slows the instability of Ellis--Bronnikov wormholes}

\author{Nikita M. Shirokov}
\email{shirokov.nm@phystech.edu}
\affiliation{\textit{Moscow Institute of Physics and Technology, Moscow, Russia}}
\author{Ilya Nachevsky}
\email{ommnammashivaya@gmail.com}
\affiliation{\textit{Higher School of Economics, Moscow, Russia}}
\date{\today}

\begin{abstract}
Rotation weakens the radial instability of Ellis--Bronnikov wormholes in slow-rotation perturbation theory. We investigate its nonlinear effect by evolving stationary rotating drainholes supported by a phantom scalar field. We construct the backgrounds with an independent spectral solver, validate them against published symmetric solutions, and identify a misprint in a published field equation that did not affect the original authors' numerical results. Evolutions perturbed only by discretization error show that rotation slows the collapse. At intermediate spins some throats expand, with the direction of departure depending on the effective perturbation. At the fastest evolved spin, the throat's areal radius shows no growing departure through evolution time $t=150$ in code units, about 28 e-folding times of the static instability, at three resolutions. Two other rapidly rotating models also show no growth over their reliable evolution intervals. Collapsing rotating throats emit an axisymmetric quadrupolar gravitational-wave burst whose energy increases between the two spins compared. Second-order initial data excite the instability through their background truncation error. The results establish slower collapse and prolonged absence of detected radial growth for the sampled asymmetric backgrounds, over finite times from initially axisymmetric data.
\end{abstract}

\maketitle

\section{Introduction}
\label{sec:intro}

The Ellis--Bronnikov wormhole, a throat held open by a phantom scalar field~\cite{ellis1973,bronnikov1973,morris_thorne1988,visser1995}, is unstable: depending on the sign of the perturbation it collapses to a black hole or expands without bound, on a timescale set by the throat radius~\cite{gonzalez2009lin,gonzalez2009,shinkai2002,bronnikov2011,cremona2019}. Nonlinear 3+1 evolutions confirm both branches for the isolated throat~\cite{shirokov2026} and carry them into binaries, in which every configuration ends in collapse, expansion or a merged horizon within a few dynamical times~\cite{shirokov2026c}. A persistent traversable wormhole needs a mechanism that removes this instability.

Rotation is the oldest candidate~\cite{matos2006}. Spinning Ellis--Bronnikov solutions are known perturbatively~\cite{kashargin2008} and, beyond perturbation theory, from the elliptic construction of Kleihaus, Kunz and Chew~\cite{kleihaus_kunz2014,chew2016}, including symmetric and asymmetric families: a sequence that starts at the static throat and ends, at equatorial rotation velocity $v_e=1$, on an extremal Kerr black hole, the phantom field vanishing in the limit; no closed form exists~\cite{volkov2021}.

On the stability side, Azad, Bl\'azquez-Salcedo, Khoo and Kunz~\cite{azad2023,azad2024} carried the radial mode to second order in spin: the static unstable mode weakens, with a positive $O(J^2)$ correction for every asymmetry, while for nearly symmetric wormholes a second unstable mode grows out of a static zero mode. The illustrated mode crossings near $30\%$ of the extremal value concern nearly symmetric wormholes. The authors conjecture that the modes merge beyond the reliable range of the quadratic approximation and the instability disappears, as in five dimensions~\cite{dzhunushaliev2013}. Quasinormal modes of the rapidly rotating family have also been computed~\cite{khoo2024}, and a later slow-rotation study derives the equations for quadrupole perturbations~\cite{azad2025quad}. In the charged analogue the merged modes continue as a complex pair (oscillating growth), and the growth rate reaches zero only in the extremal limit~\cite{blazquez2025}. A perturbative method cannot decide which continuation rotation takes in four dimensions, and, to our knowledge, no rotating wormhole has yet been evolved nonlinearly.

This paper builds the tools for that evolution and applies them. We construct two families of backgrounds as exact stationary solutions, distinguished by the asymmetry $\gamma$ between their two ends (defined in Sec.~\ref{sec:model}). The symmetric family ($\gamma=0$), massless when static, is the one with published data, the curves of Ref.~\cite{kleihaus_kunz2014} and the scalar charges of Ref.~\cite{chew2016}, and we use it to validate the solver and the 3D initial data across the spin range. The drainhole family ($\gamma=-\pi$), whose static member is the mass-1 drainhole evolved in Refs.~\cite{shirokov2026,shirokov2026c}, lies within the asymmetric solution space studied in Ref.~\cite{chew2016}; it is the family we construct independently and evolve.

The solver tests every solution against identities it never uses. These checks exposed a misprint in a published field equation (Sec.~\ref{sec:model:eq12}) and showed that second-order initial data trigger the instability at their own error level (Sec.~\ref{sec:slow}). We then put the backgrounds on the adaptive-mesh-refinement (AMR) evolution grid, verify them at fourth order, and evolve the drainhole at nine spins (Table~\ref{tab:ladder}). We find that rotation slows the collapse, as second-order perturbation theory predicts. Some intermediate-spin evolutions expand, with the effective perturbation determining the direction of departure. At the three sampled higher spins the throat shows no growing areal-radius departure within the reliable intervals; the fastest evolved model remains close to its stationary radius throughout the three-resolution study.

Section~\ref{sec:model} gives the family and the corrected field equations; Sec.~\ref{sec:solver} the solver, the validation against the published data, and the map of the parameter space (Table~\ref{tab:backgrounds}); Sec.~\ref{sec:setup} the numerical setup and the runs (Table~\ref{tab:ladder}); Sec.~\ref{sec:slow} the evolutions; Sec.~\ref{sec:conclusions} the conclusions. Appendix~\ref{app:conv} gives the convergence of the evolutions of the slowest- and fastest-spinning models; Appendix~\ref{app:interface} the choice of timestep. We use $G=c=1$; the evolutions use the code units of Refs.~\cite{shirokov2026,shirokov2026c}, with throat parameter $\eta_0=2$ (defined in Sec.~\ref{sec:model}), in which the static drainhole has mass $M=1$.

\section{The spinning wormhole family}
\label{sec:model}

The matter model is that of our previous work~\cite{shirokov2026,shirokov2026c}: Einstein gravity coupled to a phantom scalar field $\psi$,
\begin{equation}
S=\frac{1}{16\pi}\int d^4x\,\sqrt{-g}\left(R+2\,g^{\mu\nu}\partial_\mu\psi\,\partial_\nu\psi\right),
\label{eq:action}
\end{equation}
whose kinetic term has the reversed sign that makes $\psi$ a ghost. This sign lets the scalar's stress tensor violate the null energy condition, which a static traversable throat requires~\cite{morris_thorne1988,visser1995}. Varying the action with respect to the metric and to $\psi$ gives the field equations of the model [Eqs.~(2)--(3) of Ref.~\cite{chew2016}],
\begin{equation}
R_{\mu\nu}=-2\,\partial_\mu\psi\,\partial_\nu\psi,\qquad
\partial_\mu\!\left(\sqrt{-g}\,g^{\mu\nu}\partial_\nu\psi\right)=0:
\label{eq:fieldeqs}
\end{equation}
Einstein's equations with the phantom's stress tensor as source, and the wave equation for the phantom on the curved background. Spinning the wormhole of Refs.~\cite{shirokov2026,shirokov2026c} up requires the stationary rotating solution of the same theory, constructed numerically inside an ansatz with stationarity and axisymmetry built in. We take the ansatz of Refs.~\cite{kleihaus_kunz2014,chew2016}, who first constructed these solutions,
\begin{equation}
\begin{split}
ds^2&=-e^{f}dt^2+p^2e^{-f}\Big[e^{\nu}\left(d\eta^2+h\,d\theta^2\right)\\
&\quad+h\sin^2\!\theta\left(d\varphi-\omega\,dt\right)^2\Big],
\end{split}
\label{eq:metric}
\end{equation}
with $\eta\in(-\infty,\infty)$ and four unknown functions $f$, $p$, $\nu$, $\omega$ of $\eta$ and $\theta$ [Eq.~(4) of Ref.~\cite{chew2016}]. The field equation for $p$ decouples from the others, and its solution with asymptotically flat boundary conditions is $p=1$ [Eqs.~(5)--(6) of Ref.~\cite{chew2016}]; we set $p=1$ from here on.

In the trivial limit $f=\nu=\omega=0$ the spatial metric is $d\eta^2+h\left(d\theta^2+\sin^2\!\theta\,d\varphi^2\right)$ with the fixed $h=\eta^2+\eta_0^2$: $\sqrt{h}$, the radius of the sphere at coordinate $\eta$, never vanishes; its minimum, $\eta_0$ at $\eta=0$, is the throat, the minimal sphere joining the two asymptotically flat ends $\eta\to\pm\infty$, so the spacetime has no center. This choice of $h$ builds the two-ended, horizonless topology into Eq.~\eqref{eq:metric}; conversely, $\eta_0\to0$ closes the throat and lets the same ansatz hold a black hole, which is how extremal Kerr enters in Sec.~\ref{sec:model:eq12}.

Our previous wormhole sits inside this form: with $\omega=\nu=0$ and the closed-form $\arctan$ profile for $f$~\cite{chew2016}, Eq.~\eqref{eq:metric} is the static drainhole of Refs.~\cite{shirokov2026,shirokov2026c}, with $\eta_0=a$ its throat parameter and $\eta$ the isotropic radius used there, unfolded through the throat (Sec.~\ref{sec:setup:id}); $f=0$ gives the massless Ellis wormhole.

Substituting the ansatz~\eqref{eq:metric} into the field equations~\eqref{eq:fieldeqs} reduces them in three steps. We follow the reduction of Kleihaus and Kunz~\cite{kleihaus_kunz2014}, given in full in Sec.~II of Ref.~\cite{chew2016}; equation numbers in brackets below refer to Ref.~\cite{chew2016}. First, the phantom field equation (the wave equation in Eq.~\eqref{eq:fieldeqs}, which stationarity and axisymmetry reduce to a function of $\eta$ alone) integrates in closed form to $\psi=(D/\eta_0)\left[\arctan(\eta/\eta_0)-\pi/2\right]$ [Eqs.~(7)--(9)]: the scalar enters the rest of the problem only through its constant of integration $D$, the scalar charge. Second, because $\partial_\mu\psi$ points along $\eta$, the source term $-2\,\partial_\mu\psi\,\partial_\nu\psi$ survives only in the $\eta\eta$ Einstein equation; each other nontrivial component ($tt$, $t\varphi$, $\theta\theta$, $\varphi\varphi$, $\eta\theta$) is a statement about the metric functions alone. Two of them are second-order elliptic equations,
\begin{align}
0={}&\partial_\eta\!\left(h s\,\partial_\eta f\right)+\partial_\theta\!\left(s\,\partial_\theta f\right)\nonumber\\
&-h s^3 e^{-2f}\!\left[h(\partial_\eta\omega)^2+(\partial_\theta\omega)^2\right],
\label{eq:feq}\\
0={}&\partial_\eta\!\left(h^2 s^3 e^{-2f}\partial_\eta\omega\right)\nonumber\\
&+\partial_\theta\!\left(h s^3 e^{-2f}\partial_\theta\omega\right),
\label{eq:omeq}
\end{align}
with $s\equiv\sin\theta$ [Eqs.~(10)--(11)]: Eq.~\eqref{eq:feq} is a Laplace-type equation for the redshift $f$, sourced by the squared gradients of the frame dragging (rotation gravitates), and Eq.~\eqref{eq:omeq} is a total divergence, angular-momentum conservation, forcing the same $J$ through every $\eta=\mathrm{const}$ surface. The single off-diagonal component is only first order in $\nu$,
\begin{equation}
\begin{split}
0&=-h s\,\partial_\eta f\,\partial_\theta f+h\cos\theta\,\partial_\eta\nu\\
&\quad+\eta s\,\partial_\theta\nu+h^2 s^3 e^{-2f}\partial_\eta\omega\,\partial_\theta\omega,
\end{split}
\label{eq:constraint}
\end{equation}
so $\nu$ follows by integration once $f$ and $\omega$ are known [Eq.~(13)]. (The remaining diagonal components give a second-order equation for $\nu$, Sec.~\ref{sec:model:eq12}.) Third, the one sourced component, $R_{\eta\eta}=-2D^2/h^2$ [Eq.~(18)], rearranges into an algebraic identity for the charge [Eq.~(19)]: the metric functions fix $D^2$ at every point, so its constancy over the domain is an independent test of the solution. The system for three unknowns is overdetermined. The Bianchi identities make it consistent on an exact solution, so a numerical scheme must choose which subset to solve, a choice Sec.~\ref{sec:model:eq12} shows matters. The boundary conditions select the family: Minkowski at $\eta\to+\infty$, and $f\to\gamma$, $\omega\to\omega_\infty$ at $\eta\to-\infty$, so the constant $\gamma$ measures how unequal the two ends are and $\omega_\infty$ sets the spin.

A solution of this system is three functions and one number, $(f,\nu,\omega;D)$. Every physical quantity we quote (the columns of Table~\ref{tab:backgrounds}) is read off them directly~\cite{kleihaus_kunz2014,chew2016}:
\begin{itemize}
\item far from the throat the metric settles to the field of a point mass, $f\to-2\mu_+/\eta$ as $\eta\to+\infty$ and $f\to\gamma+2\mu_-/\eta$ as $\eta\to-\infty$, so the end masses $\mu_\pm$ are the $1/\eta$ coefficients of $f$;
\item likewise $\omega\to2J/\eta^3$, and the charges obey the Smarr-type relation $\mu_++\mu_-=2\,\omega_\infty J$~\cite{chew2016};
\item the throat is the sphere of minimal area, at $\eta_t=0$ when the two ends are equal ($\gamma=0$) and displaced toward the positive-mass end otherwise ($\eta_t=\eta_0/2$ for the static drainhole): its equatorial circumference gives the radius $R_e=e^{-f/2}\sqrt{h}\,\big|_{\eta_t,\,\theta=\pi/2}$, its meridians the polar radius $R_p$, its area the areal radius $R_A$; the three coincide while the throat is round and split as rotation flattens it;
\item the throat is dragged at the angular velocity $\Omega=\omega\,\big|_{\eta_t,\,\theta=\pi/2}$, and $v_e=R_e\,\Omega$ is its equatorial linear velocity in units of $c$: the dimensionless spin label used throughout, 0 when static and 1 at the extremal Kerr endpoint.
\end{itemize}

We construct two one-parameter families, labelled by $\omega_\infty$ (equivalently $J$ or $v_e$):
\begin{itemize}
\item[(a)] \emph{the drainhole family}, $\eta_0=2$, $\gamma=-\pi$: its static member, with end masses $\mu_\pm=\pm1$, is the wormhole of Refs.~\cite{shirokov2026,shirokov2026c}; its end-mass asymmetry is $C=1$ in the notation of Refs.~\cite{azad2023,azad2024}.
\item[(b)] \emph{the symmetric family}, $\gamma=0$ ($C=0$), that of Refs.~\cite{kleihaus_kunz2014,chew2016}: equal masses on both ends, the massless Ellis wormhole when static; for the 3D checks it is rescaled $\eta_0\,{=}\,1\to2$, so its throat sits on family (a)'s grid.
\end{itemize}

\subsection{A misprint in the published $\nu$ equation}
\label{sec:model:eq12}

Refs.~\cite{kleihaus_kunz2014,chew2016} close the system with a second-order equation for $\nu$, Eq.~(12) of Ref.~\cite{chew2016}. The printed equation contains a misprint. Chew, Kleihaus and Kunz have confirmed that their original numerical programs used the correct equation, so the misprint does not affect their numerical solutions or results (personal communication). The scalar-free component $R_{\theta\theta}=0$, with the second derivatives of $f$ and $\omega$ eliminated by Eqs.~\eqref{eq:feq}--\eqref{eq:omeq}, gives
\begin{equation}
\begin{split}
0&=\partial_\eta\!\left(h s\,\partial_\eta\nu\right)+s\,\partial_\theta^2\nu-\cos\theta\,\partial_\theta\nu+s\,(\partial_\theta f)^2\\
&\quad-h s^3 e^{-2f}\left[h(\partial_\eta\omega)^2+2(\partial_\theta\omega)^2\right],
\end{split}
\label{eq:nueq}
\end{equation}
where the printed form has $-2(\partial_\theta\omega)^2$ and no $s\,(\partial_\theta f)^2$ term. Extremal Kerr fits the ansatz~\eqref{eq:metric} in closed form (with $\eta_0=0$), and substituted symbolically, at 160-digit arithmetic, it satisfies Eqs.~\eqref{eq:feq}--\eqref{eq:constraint}, Eq.~\eqref{eq:nueq} and the $D^2$ identity exactly, while leaving a residual of order $0.1$--$0.4$ near the horizon (at $M=1$) in the printed $\nu$ equation alone. The error also appears numerically: solutions of the printed elliptic system converge at every resolution, yet violate Eq.~\eqref{eq:constraint} and the $D^2$ identity, which that system never uses, by an amount that scales as $J^4$ and does not shrink with the grid, the signature of a difference in the continuous equations rather than in the numerics. Our solver uses neither $\nu$ equation (Sec.~\ref{sec:solver}), and Eq.~\eqref{eq:nueq} holds on all 200 saved solutions of both families to $\le7\times10^{-8}$ of its largest term within $20\,\eta_0$ of the throat, where the printed form differs by up to $53\%$. The agreement with their published solutions provides a separate validation of our implementation (Sec.~\ref{sec:solver}).

\section{Backgrounds across the spin range}
\label{sec:solver}

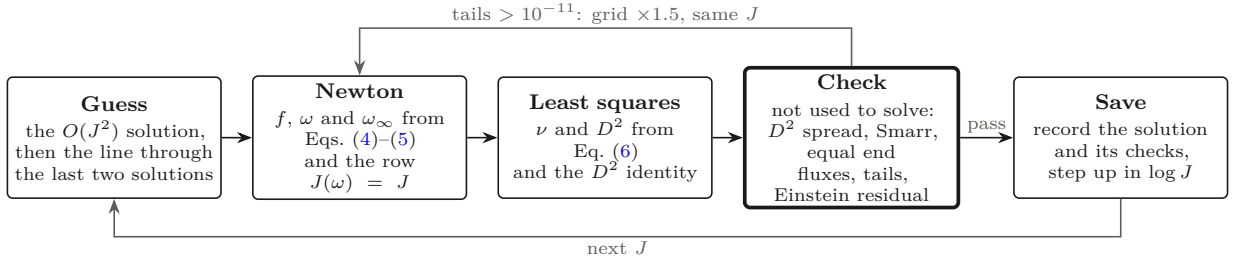
\begin{figure*}[t]
\centering
\begin{tikzpicture}[
  font=\scriptsize, node distance=4mm,
  box/.style={draw=figink, line width=0.6pt, rounded corners=2pt, align=center,
              text width=26.5mm, minimum height=16mm, inner sep=2.5pt, text=figink},
  gate/.style={box, line width=1.4pt},
  flow/.style={-{Stealth[length=2mm]}, line width=0.6pt, draw=figink},
  back/.style={-{Stealth[length=2mm]}, line width=0.6pt, draw=figmuted},
  tag/.style={text=figmuted, inner sep=1.5pt}]
\node[box] (seed) {{\footnotesize\bfseries Guess}\\[2pt]
  the $O(J^2)$ solution,\\ then the line through\\ the last two solutions};
\node[box, right=of seed] (newton) {{\footnotesize\bfseries Newton}\\[2pt]
  $f$, $\omega$ and $\omega_\infty$ from\\ Eqs.~(\ref{eq:feq})--(\ref{eq:omeq})\\ and the row $J(\omega)=J$};
\node[box, right=of newton] (nu) {{\footnotesize\bfseries Least squares}\\[2pt]
  $\nu$ and $D^2$ from\\ Eq.~(\ref{eq:constraint})\\ and the $D^2$ identity};
\node[gate, right=of nu] (gates) {{\footnotesize\bfseries Check}\\[2pt]
  not used to solve:\\ $D^2$ spread, Smarr,\\ equal end fluxes, tails,\\ Einstein residual};
\node[box, right=7mm of gates] (out) {{\footnotesize\bfseries Save}\\[2pt]
  record the solution\\ and its checks,\\ step up in $\log J$};
\draw[flow] (seed) -- (newton);
\draw[flow] (newton) -- (nu);
\draw[flow] (nu) -- (gates);
\draw[flow] (gates) -- node[tag, above] {pass} (out);
\draw[back] (gates.north) -- ++(0,5mm) -| node[tag, pos=0.25, above] {tails $>10^{-11}$: grid $\times1.5$, same $J$} (newton.north);
\draw[back] (out.south) -- ++(0,-5mm) -| node[tag, pos=0.25, below] {next $J$} (seed.south);
\end{tikzpicture}
\caption{The solver loop for one family (Sec.~\ref{sec:solver}). Thin boxes solve equations; the heavy box tests identities the solver never uses. A tail above $10^{-11}$ refines the grid and solves the same $J$ again; each solution is the initial guess for the next $J$. The loop gave 462 solutions in 20 continuation sweeps: 200 on the two families of Sec.~\ref{sec:model}, both reaching $v_e\approx0.99$, and the rest on the comparison families of Table~\ref{tab:anchors}.}
\label{fig:solver}
\end{figure*}

\begin{figure}[t]
\centering
\includegraphics[width=\columnwidth]{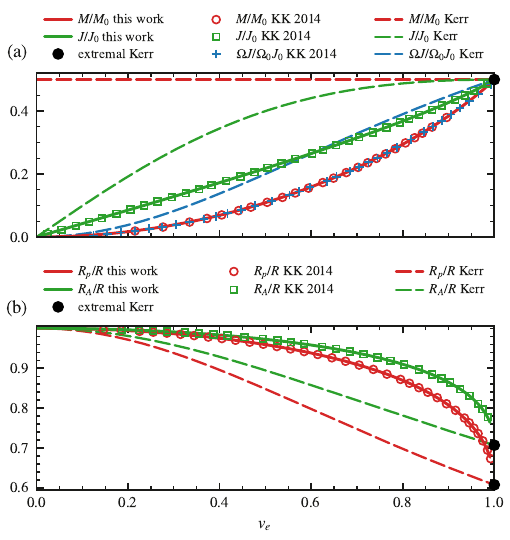}
\caption{The symmetric family ($\eta_0=1$, $\gamma=0$; 81 solutions, $v_e=0.02$--$0.991$, plus the exact static member) in the normalization of Ref.~\cite{kleihaus_kunz2014} ($M_0=R_e$, $J_0=R_e^2/2$, $\Omega_0=1/R_e$): our curves, the digitized points of their Fig.~1 (open symbols), and closed-form Kerr (dashed). The digitization's reading error, measured by reading their printed Kerr curves the same way, is rms $1$--$6\times10^{-4}$; digitized minus ours is rms $2.5$--$8\times10^{-4}$, max $2.5\times10^{-3}$, over all five curves. The two solvers agree to the precision a printed figure carries.}
\label{fig:kk2014}
\end{figure}

\begin{figure*}[t]
\centering
\includegraphics[width=\textwidth]{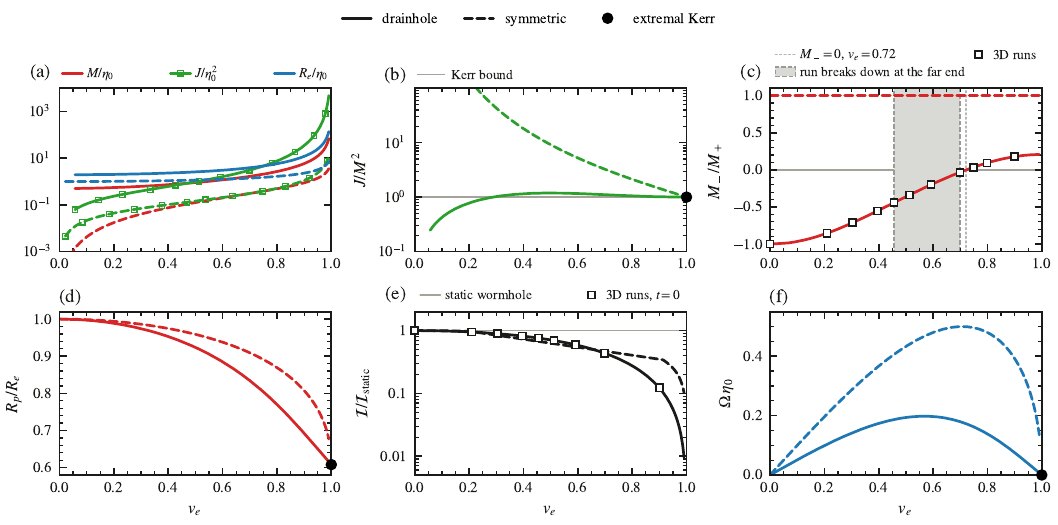}
\caption{The drainhole ($\eta_0=2$, $\gamma=-\pi$) and symmetric ($\eta_0=1$, $\gamma=0$) families across the spin range, against $v_e$: (a) the mass $M=\mu_+$ of the $\eta\to+\infty$ end, the angular momentum and the equatorial throat radius; (b) $J/M^2$ against the Kerr bound; (c) the end-mass ratio $M_-/M_+=\mu_-/\mu_+$, which changes sign along the drainhole family at $v_e=0.72$ (dotted), with the backgrounds of the 3D runs (squares); in the shaded band ($v_e=0.46$--$0.70$) the far end has negative mass and the throat does not collapse, and the 3D evolution breaks down before $t=150$ at the puncture $r=0$ into which the far end is compactified; (d) the flattening $R_p/R_e$; (e) the exotic matter: the phantom field's integrated null-energy violation over the $t=\,$const slice per equatorial throat radius, $\mathcal{I}=\int(\rho+p_\eta)\,dV/R_e$~\cite{visser2003} ($\rho$: energy density, $p_\eta$: radial pressure), relative to the static member's $\mathcal{I}_{\rm static}$, with 3D runs (squares) placed by the scalar charge measured on each run's own $t=0$ slice, which reproduces its background's to $3\times10^{-4}$; (f) the throat angular velocity, which peaks and falls to zero at extremality. Both families end on extremal Kerr, and the faster they spin, the less exotic matter each unit of throat radius needs.}
\label{fig:trends}
\end{figure*}

We solve the system of Sec.~\ref{sec:model} in five steps (Fig.~\ref{fig:solver}):
\begin{enumerate}
\item\emph{Compactify.} The coordinate $x=(2/\pi)\arctan(\eta/\eta_0)$ maps both asymptotic ends to $x=\pm1$, and the unknowns $f$ and $\omega$ are expanded in Chebyshev polynomials of $x$ and even cosine modes of $\theta$. The fastest-spinning solutions use a log-stretched variant of $x$.
\item\emph{Solve.} The two elliptic equations~\eqref{eq:feq}--\eqref{eq:omeq} are solved for the expansion coefficients by Newton iteration (guess, linearize, correct) until their residual sits at round-off.
\item\emph{Recover $\nu$ and $D$.} The first-order constraint~\eqref{eq:constraint} and the $D^2$ identity together overdetermine $\nu$ and the single number $D^2$, so both are extracted as one linear least-squares problem, with $\nu\to0$ at both ends and without any second-order $\nu$ equation; the leftover misfit measures the error.
\item\emph{Continue in spin.} Every converged solution is the initial guess for the Newton iteration at a slightly higher spin, with $J$ as the continuation parameter ($\omega_\infty$ is not monotone along a family), and the grid refines by a factor $1.5$, from $48\times12$ up to $192\times64$, whenever the asymptotic-tail check of step 5 fails.
\item\emph{Check.} Every solution is tested against five identities the solver never uses (Table~\ref{tab:gates}); the asymptotic tails are the one check the loop acts on (step 4).
\end{enumerate}
The fifth identity, a symbolically generated residual of the full Einstein equations evaluated within $20\,\eta_0$ of the throat, is the sharpest. It is $9\times10^{-11}$ at $J=0.25$ and $1.5\times10^{-9}$ at $J=1$, against the static member's $1.4\times10^{-11}$ round-off floor; the $O(J^2)$ perturbative data give $6\times10^{-5}$ and $2\times10^{-2}$ on meshes of their own. It stays at a few $10^{-9}$ up to $v_e\approx0.85$ and reaches $2\times10^{-6}$ at the fastest spins. Farther than $20\,\eta_0$ the residual's normalization inflates round-off and the test is switched off; the outer region is covered by the asymptotic-tail check and the 3D constraint checks of Sec.~\ref{sec:setup:checks}. The $J=0.25$ solution solved on two grids and two radial maps agrees in its global charges to $6\times10^{-7}$ or better, and each of the ten backgrounds evolved in Sec.~\ref{sec:slow} is exported as a table of $2049\times129$ to $8193\times513$ points, whose bilinear error in $f$ is $\le1.5\times10^{-7}$ up to $v_e=0.8$ and $6\times10^{-7}$ at $v_e=0.9$.

\begin{table}[t]
\caption{The identity checks of step 5: five relations the solver never uses, and the largest relative violation over the 200 solutions of the two families. The asymptotic tails drive the refinement of step 4. The Einstein residual is evaluated within $20\,\eta_0$ of the throat; its spin dependence is given in the text.}
\label{tab:gates}
\scriptsize
\begin{ruledtabular}
\begin{tabular}{ll}
Identity & Largest violation \\
\hline
Spread of $D^2$ over the domain & $1.5\times10^{-7}$ \\
Smarr relation $\mu_++\mu_-=2\,\omega_\infty J$ & $1.3\times10^{-6}$ \\
Equal $J$ flux through the two ends & $4\times10^{-9}$ \\
Asymptotic tails (refinement trigger) & $10^{-11}$ \\
Einstein residual & $1.4{\times}10^{-11}$--$2.2{\times}10^{-6}$ \\
\end{tabular}
\end{ruledtabular}
\end{table}

\begin{table}[t]
\caption{Scalar charge $D$ of the comparison solutions ($\eta_0=1$, $\omega_\infty=0.5$) against Ref.~\cite{chew2016}. Along each fixed-$\gamma$ family $\omega_\infty$ first rises with $J$, peaks and then falls, so $\omega_\infty=0.5$ is reached twice, by a slower and a faster solution; the published values are the faster ones. Our values are read along each family at $\omega_\infty=0.5$, interpolating between the two neighbouring solutions; the $\gamma=2$ family continued to $J=172$ confirms no third crossing exists. Both crossings are included in our data release.}
\label{tab:anchors}
\scriptsize
\begin{ruledtabular}
\begin{tabular}{llll}
$\gamma$ & $D$ (this work) & $D$~\cite{chew2016} & $D$, slower solution \\
\hline
$-1$ & 0.716385 & 0.71636 & 0.969193 \\
$0$  & 0.552771 & 0.55277 & 0.986013 \\
$1$  & 0.431465 & 0.43155 & 1.008136 \\
$2$  & 0.337598 & 0.33984 & 1.048161 \\
\end{tabular}
\end{ruledtabular}
\end{table}

Figure~\ref{fig:kk2014} validates the symmetric family against the published curves. Of the four solutions tabulated in Ref.~\cite{chew2016} at $\omega_\infty=0.5$, two ($\gamma=0,-1$) agree to all printed digits and two ($\gamma=1,2$) differ from ours by $2\times10^{-4}$ and $6.6\times10^{-3}$ of their value, consistent with their finite-difference error growing with $\gamma$ (Table~\ref{tab:anchors}).

\begin{table*}[t]
\caption{The two families at representative spins, in units of $\eta_0$: rotation velocity $v_e$, angular momentum, mass of the $\eta\to+\infty$ end, ratio of end masses, scalar charge, equatorial throat radius, flattening, and throat angular velocity---each read off the solution's functions as defined in Sec.~\ref{sec:model}. Every row is a converged solution, tested against the identities of Table~\ref{tab:gates}.}
\label{tab:backgrounds}
\scriptsize
\begin{ruledtabular}
\begin{tabular}{llllllllll}
family & $v_e$ & $J/\eta_0^2$ & $\mu_+/\eta_0$ & $\mu_-/\mu_+$ & $D/\eta_0$ & $R_e/\eta_0$ & $R_p/R_e$ & $\Omega\,\eta_0$ \\
\hline
(a) drainhole & 0.213 & 0.2535 & 0.5574 & $-0.850$ & 1.1004 & 2.0319 & 0.9873 & 0.1046 \\
              & 0.513 & 1.0278 & 0.9302 & $-0.339$ & 1.0006 & 2.6271 & 0.9194 & 0.1952 \\
              & 0.699 & 3.0417 & 1.6557 & $-0.032$ & 0.8598 & 3.8864 & 0.8351 & 0.1798 \\
              & 0.900 & 34.550 & 5.8400 & $+0.179$ & 0.5327 & 11.911 & 0.6926 & 0.0755 \\
              & 0.950 & 152.16 & 12.316 & $+0.201$ & 0.3785 & 24.751 & 0.6504 & 0.0384 \\
              & 0.990 & 4053.8 & 63.666 & $+0.208$ & 0.1698 & 127.36 & 0.6165 & 0.0078 \\
\hline
(b) symmetric & 0.239 & 0.0542 & 0.0252 & 1 & 0.9881 & 1.0299 & 0.9913 & 0.2323 \\
              & 0.434 & 0.1158 & 0.0906 & 1 & 0.9583 & 1.1101 & 0.9699 & 0.3911 \\
              & 0.667 & 0.2683 & 0.2666 & 1 & 0.8854 & 1.3420 & 0.9197 & 0.4969 \\
              & 0.847 & 0.7007 & 0.6305 & 1 & 0.7653 & 1.8836 & 0.8440 & 0.4499 \\
              & 0.898 & 1.0989 & 0.8685 & 1 & 0.7043 & 2.2722 & 0.8092 & 0.3952 \\
              & 0.952 & 2.4701 & 1.4383 & 1 & 0.5971 & 3.2703 & 0.7539 & 0.2911 \\
              & 0.990 & 12.235 & 3.4312 & 1 & 0.4157 & 7.0595 & 0.6795 & 0.1402 \\
\end{tabular}
\end{ruledtabular}
\end{table*}

Figure~\ref{fig:trends} and Table~\ref{tab:backgrounds} map our two sequences within the rotating solution space studied in Refs.~\cite{kleihaus_kunz2014,chew2016}. The drainhole sequence supplies the backgrounds for the nonlinear evolutions below. Two features of these sequences matter later. Both families violate the Kerr bound (Fig.~\ref{fig:trends}b): the symmetric family at every spin, with $J/\mu_+^2\approx85$ already at $v_e=0.24$, falling monotonically to 1 at extremality since its static member is massless, and the drainhole family by up to ${\sim}20\%$ at intermediate spin. A collapsing overspun throat cannot settle to a Kerr black hole of the same mass and spin, which makes these backgrounds natural initial data for tests of cosmic censorship. The drainhole's negative far-end mass $\mu_-$ rises through zero at $v_e=0.72$: above that spin both ends have positive mass. The exotic matter per throat radius falls a hundredfold along the drainhole family and tenfold along the symmetric one by $v_e=0.99$ (Fig.~\ref{fig:trends}e).

\section{Numerical setup}
\label{sec:setup}

\begin{table}[t]
\caption{The simulations: the drainhole family (a), in code units. Model D$n$ is the background with $v_e\approx n/100$ (D0: static); D21$\pm\varepsilon$ adds a radial perturbation $|\varepsilon|=10^{-3}$. $8\pi J/A$: the angular momentum as a percentage of its extremal value $A/8\pi$ ($A$: the throat area); $M_{\rm ADM}$: the mass of the $\eta\to+\infty$ end. The finest $\Delta x$ is $0.0625$, scaling as $1/N$ within a triplet. $\sigma$: the Kreiss--Oliger strength (Sec.~\ref{sec:setup:scheme}); the conformal-factor floor is $10^{-8}$ ($10^{-4}$ in D75--D90). Layer: the damping layer of Sec.~\ref{sec:setup:boundary} (off: the background-relative boundary alone). Final $t$: to two significant figures. Table~\ref{tab:results} gives the results.}
\label{tab:ladder}
\scriptsize
\setlength{\tabcolsep}{1.5pt}
\begin{ruledtabular}
\begin{tabular}{lllllllll}
model & $v_e$ & $8\pi J/A$ & $M_{\rm ADM}$ & $L$ & $N$ & $\sigma$ & layer & final $t$ \\
\hline
D0 & 0 & 0 & 1.000 & 64 & 128 & 0.1 & off & 63 \\
D21 & 0.210 & 12.3 & 1.112 & 64 & 96/128/192 & 0.1 & on & 63/100/100 \\
D21$\pm\varepsilon$ & 0.210 & 12.3 & 1.112 & 64 & 128 & 0.1 & off & 44/100 \\
D21 & 0.210 & 12.3 & 1.112 & 128 & 256 & 0.1 & off & 150 \\
D30 & 0.304 & 18.3 & 1.248 & 64 & 128 & 0.1 & off & 100 \\
D30 & 0.304 & 18.3 & 1.248 & 64 & 128 & 0.1 & off & 130 \\
D40 & 0.395 & 24.4 & 1.448 & 64 & 128 & 0.1 & off & 100 \\
D46 & 0.456 & 28.8 & 1.634 & 64 & 128 & 0.1 & off & 100 \\
D46 & 0.456 & 28.8 & 1.634 & 64 & 128 & 0.1 & off & 130 \\
D51 & 0.513 & 33.3 & 1.860 & 64 & 192 & 0.1 & on & 100 \\
D51 & 0.513 & 33.3 & 1.860 & 64 & 128 & 0.1 & off & 100 \\
D75 & 0.749 & 57.6 & 4.110 & 128 & 256 & 0.4 & on & 150 \\
D80 & 0.799 & 64.4 & 5.318 & 128 & 256 & 0.4 & on & 150 \\
D90 & 0.900 & 80.8 & 11.68 & 128 & 128/192/256 & 0.4 & on & 150 \\
\end{tabular}
\end{ruledtabular}
\end{table}

\subsection{3D initial data}
\label{sec:setup:id}

We interpolate the stationary metric~\eqref{eq:metric} onto the Cartesian AMR grid through the isotropic radius $r$, $\eta=r-\eta_0^2/4r$: the far end $\eta\to-\infty$ compactifies to the puncture $r=0$, as for the static drainhole~\cite{shirokov2026}. The spatial metric $\gamma_{ij}$ and the scalar field are read from the tables. The lapse $\alpha$ and the shift $\beta^i$ start from the stationary data, $\alpha=e^{f/2}$ and $\beta^\varphi=-\omega$, so the extrinsic curvature $K_{ij}=(D_i\beta_j+D_j\beta_i)/2\alpha$, with $D_i$ the covariant derivative of $\gamma_{ij}$, has only $r\varphi$ and $\theta\varphi$ components, and the scalar field's conjugate momentum vanishes. At the puncture the frame rotates at $\omega_\infty$.

\subsection{Validation of the interpolation onto the grid}
\label{sec:setup:checks}

We check the interpolated data in three steps:
\begin{enumerate}
\item\emph{Constraints and throat radii.} At $t=0$ the Hamiltonian and momentum constraint norms fall with resolution on every refinement level (Figs.~\ref{fig:conv02}(d,h) and~\ref{fig:conv09}(d,h) at $t=0$). In radial shells on the drainhole at $v_e=0.51$ ($L=64$, three grids), they converge at fourth order between the puncture and the throat; farther out the tables' interpolation error dominates, growing as $1/\Delta x^2$ (Hamiltonian) and $1/\Delta x$ (momentum). The throat radii along coordinate rays converge to the tables' values: to $1.4\times10^{-4}$ at $N=192$ on that drainhole and, on the symmetric family at seven spins $v_e=0.24$--$0.99$ with three resolutions each, to $1\times10^{-3}$ at worst ($N=256$).
\item\emph{Mass and angular momentum.} On the same symmetric set, the three-dimensional Arnowitt--Deser--Misner (ADM) mass integral evaluated on the grid equals the same integral on the exact table metric to $\le10^{-12}$, and, extrapolated in $1/R$ across the extraction spheres, it meets the solver's mass to at worst $0.7\%$ relative (at $v_e=0.99$); the ADM angular momentum matches to $2\times10^{-3}$ or better.
\item\emph{The static evolution.} Evolved from these data, the static drainhole reproduces the evolution of Ref.~\cite{shirokov2026c} for $t\le100$: the $t=0$ data are bit-identical, the Hamiltonian norm stays within $5\times10^{-7}$, and the collapse times agree within one time unit.
\end{enumerate}

\subsection{Evolution scheme}
\label{sec:setup:scheme}

\begin{figure}[t]
\centering
\includegraphics[width=\columnwidth]{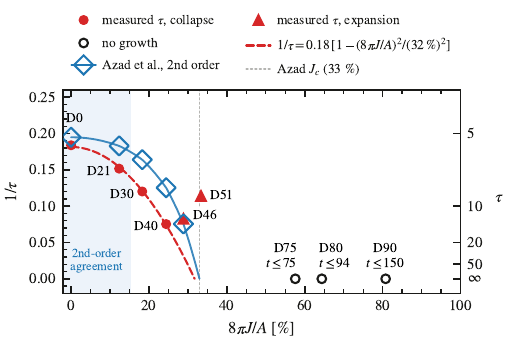}
\caption{Inverse $e$-fold time $1/\tau$ of the throat's areal radius against the angular momentum as a fraction of its extremal value (Table~\ref{tab:results}), beside the second-order rates of Refs.~\cite{azad2023,azad2024} for this family ($J_c$: the zero of their rate; the D21 point is the $N=128$ rate). Open markers at zero denote no detected growth within the stated intervals, rather than measured zero eigenvalues. The collapse rate decreases with spin; expansion markers describe the departures in the individual simulations.}
\label{fig:curve}
\end{figure}

\begin{figure*}[t]
\centering
\includegraphics[width=0.88\textwidth]{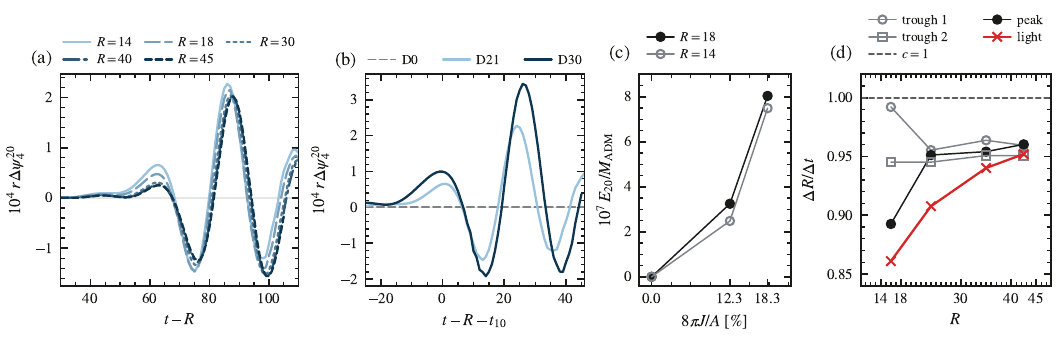}
\caption{The $(2,0)$ wave of the unperturbed rotating collapse. (a)~$r\,\Delta\psi_4^{20}$ of the D21 run on the $L=128$ box, on spheres $R=14$--45 against retarded time $t-R$. (b)~The wave against spin at $R=14$, each run on its own collapse clock $u=t-R-t_{10}$ ($t_{10}$: the time its equatorial deficit passes $10\%$); the static run (D0) is drawn at zero, its measured record below $4\times10^{-9}$. (c)~The radiated energy $E_{20}=\tfrac{1}{16\pi}\int|\int r\psi_4\,dt'|^{2}\,dt$ over the aligned burst, per unit $M_{\rm ADM}$, at $R=18$ (filled) and $R=14$ (open). (d)~The speed of each extremum between neighbouring spheres, against the local coordinate speed of light (crosses). The burst's energy grows with spin, and the static collapse radiates nothing.}
\label{fig:wave}
\end{figure*}

Evolution uses \texttt{GRTeclyn}~\cite{grteclyn,amrex,grchombo} with the setup of Refs.~\cite{shirokov2026,shirokov2026c}: the conformal and covariant Z4 (CCZ4) formulation of the Einstein equations~\cite{alic2012}, $1{+}\log$ slicing~\cite{bona1995}, Gamma-driver shift~\cite{alcubierre2003gauge}, fourth-order stencils, fourth-order Runge--Kutta time integration with $\Delta t=0.02\,\Delta x$ in every run, Kreiss--Oliger dissipation of strength $\sigma$, and a floored conformal factor at the puncture. The dissipation strength, conformal-factor floor and boundary damping differ between the slow- and fast-spin setups (Table~\ref{tab:ladder}). The larger dissipation strength suppresses a grid-scale mode that grows at the fixed mesh-refinement interfaces (Appendix~\ref{app:interface}). The larger $\sigma$ smooths the initial data more and advances the collapse by about 4 time units, so we compare rates only at equal $\sigma$.

\subsection{Long evolutions: the outer boundary}
\label{sec:setup:boundary}

Evolutions that must outlive the outer-boundary crossing time $t\approx L$ need an outer boundary that neither reflects nor drives the interior. Kreiss--Oliger dissipation cannot do it, since its strength falls as the sixth power of wavelength, so the sponge layer of Refs.~\cite{shirokov2026,shirokov2026c} is transparent to the long outgoing gauge pulse; and a flat-space Sommerfeld condition assumes asymptotics the wormhole does not have (mass 1--12 sits inside the box). Two changes in the same outer layer fix it. The Sommerfeld condition is applied to the deviation of the evolved fields from the stationary background rather than on the fields themselves; and a damping layer can in addition drive every evolved field back to that background at rate $\lambda$, $\partial_t u \to \partial_t u - \lambda\,f(r)\,(u-u_{\rm bg})$, with $f(r)$ the sponge's quartic ramp and $u_{\rm bg}$ the run's unperturbed stationary background, rebuilt exactly from the tables; $\lambda=0$ recovers the base scheme identically. Two otherwise identical D21 evolutions, damping on and off, validate the layer at $\lambda=1$: the equatorial throat radius agrees to $2\times10^{-7}$, the near-zone $\Psi_4$ signal to $10^{-3}$ of its peak, and the fine-level constraint norms are unchanged, so the layer does not touch the interior. The same comparison shows the background-relative boundary alone keeps the coarse outer levels cleaner, so the $L=64$ collapse runs from D30 to D51 take the damping off; the D21 resolution study, the D51 run at $N=192$ and the $L=128$ boxes from D75 up run with it on (Table~\ref{tab:ladder}). With this boundary the $L=64$ runs, which a standard boundary limits to the crossing time $t\approx64$, run to their final times $t=100$--$150$, with the converged span set by the constraint norms instead.

\subsection{Criteria, diagnostics and runs}
\label{sec:setup:runs}

Stability claims follow the criteria of Sec.~II of Ref.~\cite{evstafyeva2025}: a geometric measure of the growth, for which we use the areal radius defined below and the marginally outer-trapped surface (MOTS) found spectrally on every output; constraint norms that converge at the code's order while the growth proceeds; an $e$-fold time that converges to a nonzero value across resolutions; and its dependence on the perturbation, tested at D21 with added perturbations of both signs ($|\varepsilon|=10^{-3}$; Table~\ref{tab:ladder}). Every run carries three nested refinement levels on the throat, and the domain size follows the model's mass: $M_{\rm ADM}$ triples along the family (1.112 at $v_e=0.21$, 3.311 at $0.70$), so the distance from the throat to the $L=64$ boundary falls from $25\,M_{\rm ADM}$ at slow spin to under $10\,M_{\rm ADM}$ at $v_e=0.70$. Models up to $v_e=0.51$ therefore run on the $L=64$, $N=128$ grid of the static evolution of Ref.~\cite{shirokov2026c}, and models from $v_e=0.75$ on $L=128$, $N=256$ at the same finest $\Delta x$. The Newman--Penrose scalar $\Psi_4$, whose multipoles carry the outgoing gravitational wave, is extracted on spheres at $R=10$--30 ($L=64$) and $R=14$--45 ($L=128$); the signal is its $(\ell,m)=(2,0)$ mode. Studies at three resolutions test convergence at the two ends of the spin sequence (Appendix~\ref{app:conv}). Table~\ref{tab:ladder} lists every run, from the static reference through the $O(J^2)$ regime~\cite{azad2023,azad2024} to spins beyond the reach of perturbation theory.

\section{The instability across the spin range}
\label{sec:slow}

\begin{table*}[t]
\caption{Results of the simulations of Table~\ref{tab:ladder}, row by row. $t_{\rm conv}$: the time up to which the run is converged; nothing later is quoted, in the text or the figures. $t_{1\%}$: when the throat's areal radius first departs from its initial value by $1\%$. $\tau$: the $e$-fold time of the unstable mode (Sec.~\ref{sec:slow}). GW: the $(2,0)$ burst at $R=14$ arrives before $t_{\rm conv}$ (burst), is still rising at $t_{\rm conv}$ (onset), or no burst appears and the record carries only the slow drift of the near zone (none). MOTS: the marginally outer-trapped surface.}
\label{tab:results}
\scriptsize
\setlength{\tabcolsep}{3pt}
\begin{ruledtabular}
\begin{tabular}{lllllp{0.48\textwidth}}
model & $t_{\rm conv}$ & $t_{1\%}$ & $\tau$ & GW & end state \\
\hline
D0 & 60 & 43.8 & 5.4 & none & collapses; ends inside the horizon \\
D21 & 55/100/100 & 41.9/46.8/57.6 & 7.8/6.6/6.1 & burst & collapses; horizon settles at $R=2.52$ \\
D21$\pm\varepsilon$ & 42/61 & 24.2/24.2 & -- & onset & outward collapses, inward expands \\
D21 & 150 & 46.8 & -- & burst & collapses; the $(2,0)$ burst on spheres $R=14$--45 (Fig.~\ref{fig:wave}) \\
D30 & 100 & 51.0 & 8.3 & burst & collapses; horizon at $R=2.571$ at $t=100$ \\
D30 & 130 & 51.0 & 8.3 & burst & collapses; horizon settles at $R=2.567$ \\
D40 & 100 & 58.7 & 13.3 & onset & collapses; MOTS leaves the throat, $t=92$ \\
D46 & 100 & 73.7 & -- & onset & $-1.12\%$ at $t=83$, turning back by $t=100$ \\
D46 & 121 & 73.7 & 12.1 & onset & expands from $t\approx102$ \\
D51 & 100 & 94.2 & -- & none & shrinks: its effective perturbation has the opposite sign \\
D51 & 95 & 85.8 & 8.8 & none & expands from $t\approx70$ \\
D75 & 75 & -- & -- & none & no growth: areal drift $\le6\times10^{-5}$ \\
D80 & 94 & -- & -- & none & no growth: areal drift $\le7\times10^{-5}$ \\
D90 & 150 & -- & -- & none & no growth: areal drift $\le7\times10^{-5}$ at all three $N$ \\
\end{tabular}
\end{ruledtabular}
\end{table*}

\emph{Perturbative initial data.} We first evolved the second-order $O(J^2)$ backgrounds of Ref.~\cite{azad2024} at $J=0.25$ and $J=1$ on the drainhole family. Both throats expanded where the static throat collapses, on the static unstable mode with the opposite sign ($\tau=5.8$ at $J=0.25$, measured on the equatorial radius). Comparing each run's departure time with the $|\varepsilon|=10^{-3}$ perturbed runs of Table~\ref{tab:ladder} gives an effective perturbation of $\varepsilon\approx-5\times10^{-5}$ at $J=0.25$ and ${\approx}-1\times10^{-2}$ at $J=1$: growth by a factor ${\sim}200$ for a factor 4 in $J$, i.e.\ ${\sim}J^4$, the size of the $O(J^2)$ data's own truncation error, which the independent Einstein residual puts at $2\times10^{-2}$ at $J=1$ (Sec.~\ref{sec:solver}). In these evolutions, the residual of the truncated background supplies an effective perturbation. Stability studies using such data therefore require control of the background error as well as the evolution-grid error.

\emph{Measuring the growth.} The diagnostic for every evolution without an added perturbation is the areal radius $R_A(t)=\sqrt{A_{\rm sph}(t)/(4\pi)}$, where $A_{\rm sph}$ is the minimum area among the sampled coordinate spheres on a time slice. This measures the predominantly radial departure. The area of a specified surface is geometric, while the choice of coordinate spheres and the evolution time retains a dependence on the spatial coordinates and slicing. We compare this diagnostic with the circumferential radii and independently located MOTSs.

The $e$-fold time $\tau$ is the inverse slope of $\ln|dR_A/dt|$ while the departure of $R_A$ from its initial value grows from $0.2\%$ to $1\%$. We fit the rate of change rather than the departure itself because the initial transient offsets $R_A$ by a constant that differs between resolutions, and the derivative removes it. The choice of interval hardly matters: $\tau$ changes by under $6\%$ between the 0.2--1\% and 0.5--2\% intervals. Measured this way, $\tau$ self-converges across the three D21 resolutions, 7.8, 6.6 and 6.1 at $N=96$, 128 and 192, at third to fourth order ($Q=2.3$; Appendix~\ref{app:conv}); Richardson extrapolation at that measured order gives the continuum value $\tau=5.9$. This resolution study establishes convergence of the fitted timescale at D21. The zero of the fit across spins below remains an extrapolation beyond the measured collapse branch.

\emph{The collapse branch.} On the spectral backgrounds the behaviour reverses: the unperturbed throat collapses on the static mode, later and more slowly the faster it spins. Along the models D0--D40 the time at which the areal radius has fallen by $1\%$ grows steadily, and $\tau$ rises from 5.4 to 13.3 (Fig.~\ref{fig:curve}; Table~\ref{tab:results}). Each collapse ends in a black hole: the D21 horizon settles at $R=2.52$ (irreducible mass $M_{\rm irr}=R/2=1.26$), the D30 at $R=2.567$ by $t=130$; the D40's MOTS leaves the throat at $t=92$ and reads $R=3.22$, still shrinking, at $t=100$. Each evolution is perturbed only by its own discretization error; because $\tau$ comes from the growth rate, it does not depend on the perturbation amplitude, which differs between runs.

\emph{The expanding branch.} At $v_e=0.456$ ($28.8\%$ of extremal) the unperturbed throat expands instead: the areal radius falls by $1.12\%$ by $t=83$, turns, crosses zero at $t\approx102$ and grows exponentially, $\tau=12.1$. The D51 model behaves the same way ($-0.30\%$ at $t=65$--68, then exponential growth with $\tau=8.8$); its companion run, differing in resolution and boundary treatment, shrinks instead (Table~\ref{tab:results}), so at this spin the direction of departure depends on the numerical setup. The fitted expansion rate characterizes that evolution; the direction of departure alone does not identify a distinct unstable mode. No marginally outer-trapped surface forms while a throat expands, within the converged spans. Collapse and expansion are the instability's two end states, and the sign of the initial perturbation selects between them. The added radial perturbations at D21 show it directly, as for the static throat~\cite{shirokov2026c}: inward expands, outward collapses, and both depart by $1\%$ at $t\approx24$ at the same rate.

\emph{Rapid rotation.} D90 remains close to its stationary radius throughout the evolution: the areal drift stays below $7\times10^{-5}$ at all three resolutions (Appendix~\ref{app:conv}). No growing departure appears through $t=150$, about 28 $e$-folding times of the static reference in the same code units. D75 and D80 also show no growing departure over their individual reliable intervals (Table~\ref{tab:results}). Between D51 and D75 the compactified far end breaks down before the throat's behaviour can be measured (the shaded band of Fig.~\ref{fig:trends}c). There the far end has negative mass $\mu_-$ (Sec.~\ref{sec:solver}), and its puncture has no stationary state to settle into, so the evolution breaks down there, never at the throat. With $\mu_->0$, from D75 up, the runs reach $t\approx150$, the far end limiting only their converged spans (Table~\ref{tab:results}).

\emph{Against the perturbative prediction.} Figure~\ref{fig:curve} puts both branches on one axis. Where second order should hold, the two methods are close: the D21 continuum rate lies within $8\%$ of the second-order $C=1$ value of Refs.~\cite{azad2023,azad2024} ($\tau=5.9$ against 5.46), and the static member lies beside its linear mode (5.4 at $N=128$, extrapolating toward 5.1). At intermediate spin the measured mode is slower than their second-order values by more than the resolution bias ($\tau=8.3$ against 6.1 at $v_e=0.30$, 13.3 against 8.0 at 0.40), a gap that grows with spin, as expected if the neglected $O(J^4)$ terms become important. The collapse branch itself follows $a-b\,(J/A)^2$ to rms $2\times10^{-3}$ in $1/\tau$ and extrapolates to zero at $32\%$ of the extremal angular momentum, beside the zero of their second-order rate for this family ($33\%$ on this axis).

The $28.8\%$ and $33\%$ models have expanding evolutions, with fitted times $\tau=12.1$ and 8.8. These growing departures show that those simulations have not remained stationary. The second-mode crossings illustrated in Ref.~\cite{azad2024} concern $C\le0.2$, whereas our evolutions use $C=1$. The expansion therefore does not establish that their second mode becomes dominant or test their conjectured merger at the same parameters. Their $\tau$ belongs to a linear mode in Killing time and ours to a fit of the nonlinear areal radius, so the comparison at each spin is only approximate.

\emph{Gravitational waves.} The rotating collapse radiates with no nonspherical perturbation put in (Fig.~\ref{fig:wave}), whereas the static throat radiates only once a quadrupolar perturbation is added~\cite{shirokov2026c}. The signal is a single $(2,0)$ burst with ring-down period ${\approx}23$, propagating at the local light speed, its peak flat in extraction radius to $1.5\%$ over $R=30$--45 on the $L=128$ box, which reproduces the $L=64$ throat history to $3\times10^{-4}$. Aligned on each model's own collapse, the D30 burst has the D21's shape and timing at about 1.5 times the amplitude, and the radiated energy per unit mass grows as the amplitude squared, $E_{20}/M_{\rm ADM}=3.3\times10^{-7}\to8.0\times10^{-7}$. The spherical static collapse radiates nothing: its measured record at $R=14$ stays below $4\times10^{-9}$.

\section{Conclusions}
\label{sec:conclusions}

We have constructed stationary rotating Ellis--Bronnikov wormholes in two families from slow rotation to $v_e\approx0.99$, including the asymmetric drainhole sequence evolved here, and corrected a misprint in the published $\nu$ equation, Eq.~(12) of Ref.~\cite{chew2016}, to Eq.~\eqref{eq:nueq}. Additional field-equation checks exposed the misprint. The original authors confirm that their numerical implementation used the correct equation and their results are unaffected. Interpolated onto the AMR evolution grid, the solutions reproduce the solver's charges and throat radii, while the $O(J^2)$ backgrounds excite the instability through their truncation error. Quantitative evolution from these approximate backgrounds requires separate control of that error.

The nonlinear evolutions show the collapse slowing with spin and expanding departures in some intermediate-spin simulations. The radial collapse slows from $\tau=5.4$ in the static model to $13.3$ at $24\%$ of the extremal angular momentum, its rate following $a-b\,(J/A)^2$, which extrapolates to zero at $32\%$, beside the second-order prediction of $33\%$~\cite{azad2023,azad2024}. The expanding departures near $30\%$ depend on the effective perturbation, as the two D51 setups demonstrate. They establish growth in those simulations without identifying a second eigenmode.

The D90 model, at $81\%$ on this axis, shows no growing areal-radius departure through $t=150$ at three resolutions, about 28 $e$-folding times of the static reference. D75 and D80 show the same absence of detected growth over their shorter reliable intervals. These observations provide evidence of prolonged radial stability under the numerical perturbations present in the sampled high-spin evolutions.

Unlike the static collapse, the rotating collapse emits a $(2,0)$ gravitational-wave burst from unperturbed axisymmetric data, its energy growing with spin as the amplitude squared.

The evolutions end at $t\le150$, so they bound growth only over that span; a slower mode could still emerge later. Above about $29\%$ of the extremal angular momentum ($v_e\approx0.46$) the throat's equator lies inside an ergoregion. Horizonless rotating objects with an ergoregion are generically subject to a slow nonaxisymmetric instability~\cite{friedman1978,cardoso2008}; our runs, started from axisymmetric data, bound only axisymmetric growth. Only the drainhole family ($C=1$) was evolved; the symmetric family enters through the solver validation alone. The end state of each unperturbed run is set by the sign of its effective perturbation, so collapse and expansion label the sign of that perturbation, not the model. The $32\%$ zero is an extrapolation of the collapse-branch fit beyond the last collapsing model at $24\%$, not a measured endpoint.

\section{Reproducibility and cost}
\label{sec:data}

The solver, the evolution code and the figure scripts are on the \texttt{develop} branch of our \texttt{GRTeclyn} fork, archived on Zenodo~\cite{shirokov2026code}. Its directory \texttt{results/spinningwormhole} holds the solver's solutions, the headers of the background tables the evolutions start from, and every evolution's parameter file and diagnostic time series.

Every evolution ran on one NVIDIA H100 80\,GB device, with \texttt{GRTeclyn} built CUDA-native against AMReX~26.02. A run on the $L=64$, $N=128$ grid takes 7.6--11.1\,h to $t=100$; the $N=192$ members 20\,h, peaking at 53\,GiB; the $L=128$, $N=256$ runs 9.5--9.7\,h to $t=150$. The 19 evolutions of Table~\ref{tab:ladder} took 185 GPU-hours.

\begin{acknowledgments}
This work was supported by Gravity Frontiers
(\url{https://www.gravityfrontiers.org/en}), which funded the underlying
research and the development of the numerical-simulation software. The authors
thank Xiao Yan Chew, Burkhard Kleihaus and Jutta Kunz for clarifying the misprint in their published equation and its absence from their numerical implementation. They also acknowledge the GRTL Collaboration for \texttt{GRChombo}, \texttt{GRTresna},
and \texttt{GRTeclyn}.
\end{acknowledgments}

\appendix

\section{Convergence of the evolutions}
\label{app:conv}

\begin{figure*}[tp]
\centering
\includegraphics[width=0.80\textwidth]{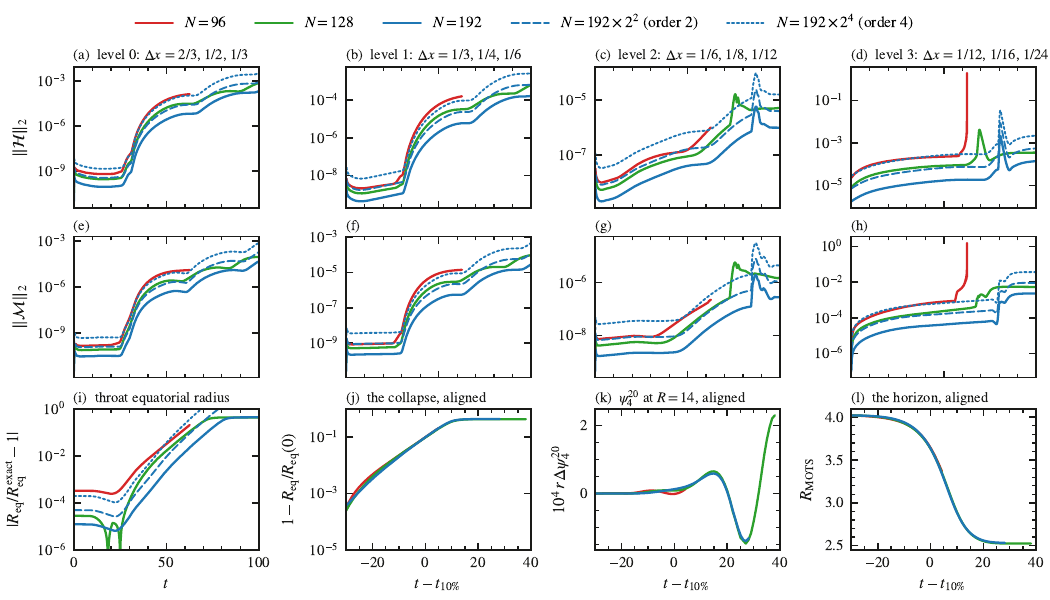}
\caption{Convergence of the D21 model over its triplet $N=96/128/192$ ($L=64$; finest $\Delta x$ in each panel title). (a)--(d)~Hamiltonian and (e)--(h)~momentum constraint norms per refinement level; dashed and dotted, the $N=192$ norms rescaled by $2^{2}$ and $2^{4}$: the $N=96$ norms lie at or above the second-order bracket on every level (above the fourth-order one late, as the coarser run collapses earlier). (i)~The equatorial throat radius against the stationary solution, falling with resolution. (j)~The collapse, each run aligned at its own $10\%$ time: the three resolutions land on one curve. (k)~$r\,\Delta\psi_4^{20}$ at $R=14$, aligned: $N=128$ and 192 overlap ($N=96$ dies in its collapse before the burst arrives). (l)~The MOTS radius, aligned: one horizon, settling at $R=2.52$. The collapse timescale self-converges at third to fourth order ($Q=2.3$); the waveform and the horizon agree across resolutions.}
\label{fig:conv02}
\end{figure*}

The two three-resolution studies are D21, the lowest nonzero spin in the collapse sequence (Fig.~\ref{fig:conv02}), and D90, the fastest evolved model (Fig.~\ref{fig:conv09}), read through the per-level norms of Appendix~\ref{app:interface}. Rescaling constraint norms between resolutions is valid because their exact value is zero. For the D21 triplet's grid spacings ($h\propto1/N$, $N=96/128/192$) the expected self-convergence factors $Q_n=(h_1^{\,n}-h_2^{\,n})/(h_2^{\,n}-h_3^{\,n})$ are $Q_3=1.95$ and $Q_4=2.69$; the fitted $e$-folding time $\tau$ of Sec.~\ref{sec:slow} gives $Q=2.3$. This factor uses differences of $\tau$, rather than its reciprocal.

\packfloatpages

\begin{figure*}[tp]
\centering
\includegraphics[width=0.80\textwidth]{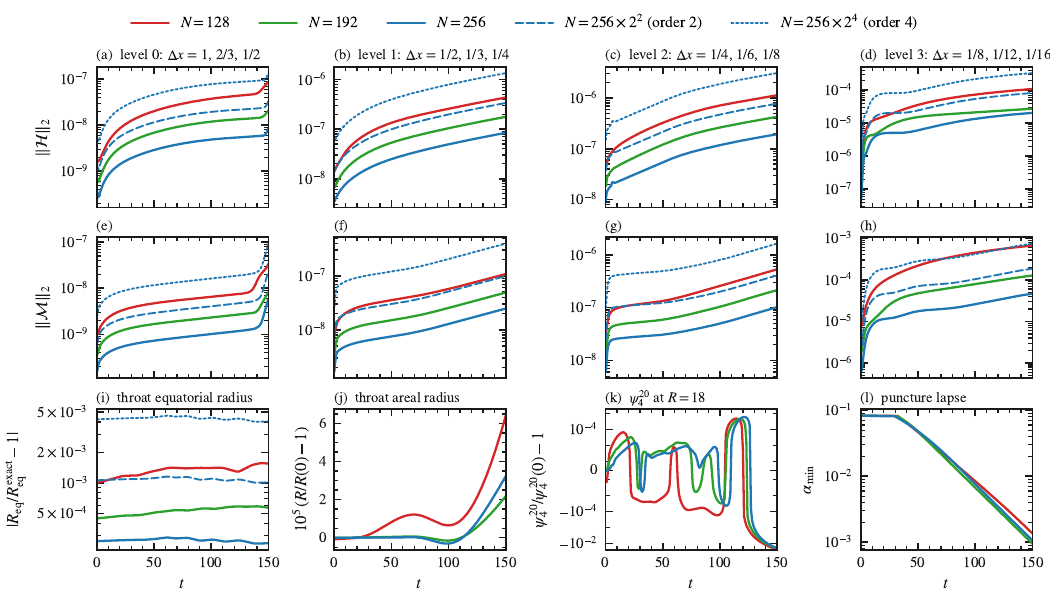}
\caption{Finite-time evolution of the D90 model over its triplet $N=128/192/256$ ($L=128$; finest $\Delta x$ in each panel title), to $t=150$. (a)--(h)~Constraint norms per level with the same second-/fourth-order brackets as Fig.~\ref{fig:conv02}: every level converges with resolution over the run. (i)~The equatorial throat radius holds the stationary value, its offset falling with resolution. (j)~The areal throat radius drifts by $\le7\times10^{-5}$ over the full run. (k)~The $(2,0)$ extraction at $R=18$, relative to its $t=0$ read-off: fluctuations $\le10^{-4}$ to $t\approx135$, a few $10^{-3}$ by $t=150$, no burst. (l)~The lapse at the $\eta\to-\infty$ puncture falls smoothly at all three resolutions (the gauge, not the geometry). The D90 throat shows no growing radius and no wave at any resolution over the run, with constraint norms converging.}
\label{fig:conv09}
\end{figure*}

\section{The refinement-interface mode and the timestep}
\label{app:interface}

The timestep $\Delta t=0.02\,\Delta x$ was adopted empirically in Refs.~\cite{shirokov2026,shirokov2026c}. We traced what sets it to a grid-scale mode pinned to the fixed coarse--fine interfaces of the nested refinement boxes, dissipation-starved at $\sigma=0.1$, whose growth rate rises with the timestep. Constraint norms evaluated per refinement level, with cells covered by finer levels masked, expose the mode by $t\approx10$, well before the domain-wide norm departs from its reference. Raising the dissipation where the mode lives removes it: at $\sigma=0.4$ the interface-level norms sit two to five orders of magnitude below the $\sigma=0.1$ reference by $t=100$. The larger timesteps this would permit went untested on the final outer-boundary setup, so all runs keep $\Delta t=0.02\,\Delta x$.

\FloatBarrier

\end{document}